\pdfoutput=1
\documentclass{ptephy_v1}

\usepackage{amsmath,amsthm,amssymb,amsfonts}
\usepackage{color}

\newcommand{\tr}{\mathrm{tr}}
\newcommand{\diag}{\mathrm{diag}}
\renewcommand{\Re}{\mathrm{Re}}

\begin{document}

\title{Non-Abelian vortex in lattice gauge theory}
\author{Arata~Yamamoto}
\affil{Department of Physics, The University of Tokyo, Tokyo 113-0033, Japan \email{arayamamoto@nt.phys.s.u-tokyo.ac.jp}}

\begin{abstract}
We perform the Monte Carlo study of the SU(3) non-Abelian Higgs model.
We discuss phase structure and non-Abelian vortices by gauge invariant operators.
External magnetic fields induce non-Abelian vortices in the color-flavor locked phase.
The spatial distribution of non-Abelian vortices suggests the repulsive vortex-vortex interaction.
\end{abstract}

\subjectindex{B01,D30}

\maketitle

\section{Introduction}

A quantum vortex is a visible topological phenomenon in the quantum world.
It is observed in superfluid helium, type-II superconductors, and atomic Bose-Einstein condensates under magnetic fields or rotation \cite{1991qvhi.book.....D,RevModPhys.82.109,RevModPhys.81.647}.
It is also believed to be the key to solve the confinement problem of the Yang-Mills theory \cite{Greensite:2011zz}.
The quantum vortex can be extended to the non-Abelian one, which has orientational moduli, in non-Abelian gauge theory. 
The non-Abelian quantum vortex is not only theoretically interesting but also phenomenologically important \cite{Tong:2005un,Eto:2006pg,Shifman:2007ce,Eto:2013hoa}.
In particular, it plays an important role in quantum chromodynamics (QCD) at high density, e.g., in compact stars \cite{Balachandran:2005ev}.

We can study the semi-classical property of vortices by solving classical equations of motion or mean-field equations.
However, this is insufficient to reveal the full quantum property of vortices. 
We need first-principle calculation of quantum field theory, such as the lattice Monte Carlo simulation.
Abelian vortices have been generated by the Monte Carlo simulation in relativistic theories \cite{Chavel:1996hh,Kajantie:1998bg,Chernodub:1998kj,Kajantie:1998zn,Kajantie:1999ih,Kajantie:2000cw,Davis:2000kv,Chernodub:2005be,Wenzel:2007uh,MacKenzie:2007ps,Bietenholz:2012ud} and non-relativistic theories \cite{PhysRevLett.75.4642,PhysRevB.72.094511,PhysRevB.89.224516,PhysRevA.95.053603}.
On the other hand, the generation of non-Abelian vortices has not been studied by the Monte Carlo simulation.

In this work, we perform the Monte Carlo study of non-Abelian vortices.
Although the most interesting theory is high-density QCD, it suffers from the fermion sign problem.
Alternatively, we adopt the non-Abelian Higgs model without the sign problem.
This model is a good starting point because it is known as an effective theory of dense quark matters \cite{Iida:2000ha,Iida:2001pg,Giannakis:2001wz}.

\section{Non-Abelian Higgs model}

Let us consider the $N$-flavor scalar field $\phi_i$ ($i=1,\cdots,N$).
Each flavor is the fundamental representation of the gauge group SU(N)$_{\rm C} \times$U(1)$_{\rm B}$, and couples to the dynamical SU(N) gauge field $A_\mu^\alpha$ and the external U(1) gauge field $A_\mu^0$.
All the flavors have the same SU(N) charge $g$ and the same U(1) charge $e$ so that the flavor symmetry is SU(N)$_{\rm F}$.
The number of colors is equal to the number of flavors.

The $d$-dimensional Euclidean action is
\begin{equation}
\label{eqScon}
S = \int d^dx \left[ \frac{1}{4} F^\alpha_{\mu\nu} F^\alpha_{\mu\nu}  + (D_\mu \phi_i)^\dagger (D_\mu \phi_i) + V[\phi] \right] 
.
\end{equation}
The potential term is
\begin{equation}
\label{eqV}
V[\phi] = -m^2\phi^\dagger_i \phi_i + \lambda (\phi^\dagger_i T^0 \phi_i)^2 +\nu (\phi^\dagger_i T^\alpha \phi_i)^2
\end{equation}
with the SU(N)$\times$U(1) generators $T^\alpha$ and $T^0$.
The covariant derivative is defined by $D_\mu= \partial_\mu + ieA_\mu^0 T^0 +igA_\mu^\alpha T^\alpha$.
The Lorentz indices $\mu,\nu$, the flavor index $i$, and the adjoint color index $\alpha$ are contracted when they are repeated.
For instance, the last term is given by $(\phi^\dagger_i T^\alpha \phi_i)^2 \equiv \sum_\alpha (\sum_i \phi^\dagger_i T^\alpha \phi_i)(\sum_j \phi^\dagger_j T^\alpha \phi_j)$.
The lattice action is 
\begin{equation}
\label{eqSlat}
\begin{split}
  S &= \sum_x \bigg[ \frac{2N}{g^2} \sum_{\mu>\nu} \left\{ 1- \frac{1}{N}\Re\tr U_{\mu\nu}(x) \right\}  + 2d \phi^\dagger_i(x) \phi_i(x)
\\
&\quad - \sum_\mu \big\{ \phi^\dagger_i(x) V_\mu(x) U_\mu(x) \phi_i(x+\hat{\mu}) 
 + \phi^\dagger_i(x+\hat{\mu}) U^\dagger_\mu(x) V^\dagger_\mu(x) \phi_i(x) \big\} + V[\phi] \bigg]
,
\end{split}
\end{equation}
where $\hat{\mu}$ is the unit lattice vector in the $x_\mu$ direction.
The summation of the Lorentz indices is explicitly written in Eq.~\eqref{eqSlat}.
The gauge fields are replaced by the dynamical SU(N) link variable $U_\mu$, and the external U(1) link variable $V_\mu$.
The field strength is replaced by the SU(N) plaquette variable $U_{\mu\nu}$.
The lattice action is always real and thus free from the sign problem.
In this and the following equations, dimensional quantities are scaled by the lattice unit.

Vortices are defined by the winding number
\begin{equation}
 Q = \frac{1}{2\pi} \oint \theta dx
\end{equation}
of the phase $\theta$ along a closed loop.
The allowed values of $Q$ depend on broken symmetry.
The global symmetry of the scalar field is SU(N)$_{\rm C}\times$SU(N)$_{\rm F}\times$U(1)$_{\rm B}/$Z$_{\rm N} \times$Z$_{\rm N}$.
The first two terms in Eq.~\eqref{eqV} spontaneously breaks U(1)$_{\rm B}$ and generates nonzero condensate.
The last term in Eq.~\eqref{eqV} favors the color-flavor locking, which breaks the color-flavor symmetry SU(N)$_{\rm C-F}$ \cite{Alford:1998mk}.
Thus, the vortices in this model are labeled by the first homotopy group $\pi_1($SU(N)$_{\rm C-F}\times$U(1)$_{\rm B}$/Z$_{\rm N}) = \mathbb{Z}$.
Non-Abelian vortices are given by the products of the U(1) and center elements, e.g., $\langle \phi_{ai} \rangle \propto e^{i\theta/N} \diag(e^{i\theta(N-1)/N},e^{-i\theta/N},\cdots,e^{-i\theta/N}) = \diag(e^{i\theta},1,\cdots,1)$.
One Abelian vortex $\langle \phi_{ai} \rangle \propto \diag(e^{i\theta},\cdots,e^{i\theta})$ is written by the superposition of $N$ non-Abelian vortices.
In this sense, non-Abelian vortices are more elemental than Abelian vortices.
For more details, see reviews \cite{Tong:2005un,Eto:2006pg,Shifman:2007ce,Eto:2013hoa}.

We focus on the $N=3$ version of this model.
This is motivated by the Ginzburg-Landau theory of dense quark matters \cite{Iida:2000ha,Iida:2001pg,Giannakis:2001wz}.
In the language of quark matters, $\phi_i$ are anti-triplet diquarks, $A_\mu^\alpha$ are dynamical gluons, and $A_\mu^0$ are external electromagnetic fields.
We note that the flavor symmetry SU(3)$_{\rm F}$ of physical diquarks is explicitly broken by the masses and electromagnetic charges of up, down, and strange quarks.
In this work, we consider the ideal model with the exact flavor symmetry.
The following analysis is possible in $d\ge 3$ dimensions.
We study the three-dimensional case for simplicity.

\section{Phase structure}

Before analyzing the topological property of this model, let us understand the basic property at vanishing external U(1) gauge fields $A_\mu^0=0$, i.e., $V_\mu=1$ for all $\mu$.
We computed by the hybrid Monte Carlo method \cite{Montvay:1994cy}.
The scalar coupling constant is $\lambda=1$ and the gauge coupling constant is $g=0$ or $g=1$.
At $g=0$, the gauge field action is dropped and all the SU(3) link variables are fixed at $U_\mu = 1$.
The lattice volume is fixed at $L^3=12^3$.
All boundary conditions are periodic.

\begin{figure}[h]
\begin{center}
 \includegraphics[width=.6\textwidth]{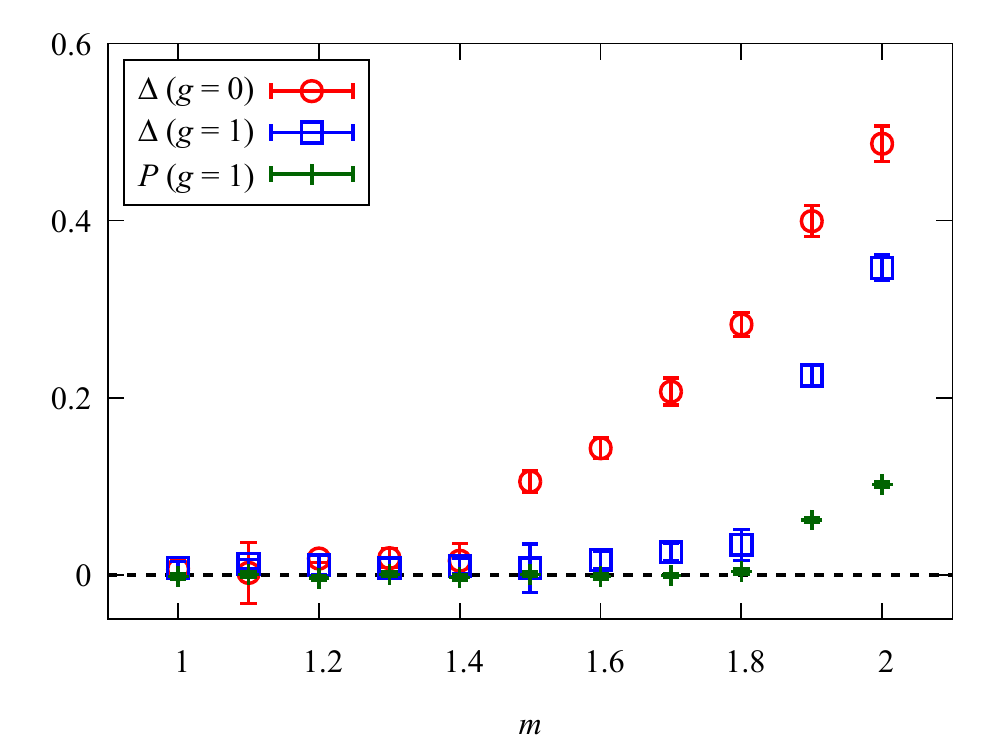}
\caption{
\label{figCM}
Condensate $\Delta$ and the Polyakov loop $P$.
The scalar coupling constant is $\nu=1$.
}
\end{center}
\end{figure}

The tachyonic mass $m$ induces the Bose-Einstein condensation.
Naively, the order parameter is the condensate $\langle \phi_i \rangle$, which is estimated by the non-local two-point function $\langle \phi^\dagger_i(x) \phi_i(y) \rangle$ in the long-range limit $|x-y| \to \infty$.
The non-local two-point function works at $g=0$ but does not work at $g \neq 0$ because it is not gauge invariant.
We consider the gauge-invariant three-particle operator
\begin{equation}
\label{eqH}
 H(x) = \frac{1}{3!} \epsilon_{ijk} \epsilon_{abc} \phi_{ai}(x) \phi_{bj}(x) \phi_{ck}(x)
,
\end{equation}
where $a$, $b$, and $c$ are the fundamental color indices \cite{Alford:2018mqj}.
We calculated the condensate of this operator
\begin{equation}
\label{eqDelta}
 \Delta = \left\{ \lim_{|x-y| \to \infty} \langle H^\dagger(x) H(y) \rangle \right\}^{1/2}
,
\end{equation}
as the order parameter.
As shown in Fig.~\ref{figCM}, the condensate is nonzero above the critical mass $m \simeq 1.4$ at $g=0$ and $m \simeq 1.8$ at $g=1$.
The critical mass is larger at $g=1$ because the scalar field gets a dressed mass by the coupling to the dynamical SU(3) gauge field.
In Fig.~\ref{figCM}, we also plot the Polyakov loop, which is defined by the dynamical SU(3) link variables as $P \equiv \langle \tr \prod_{x_3} U_3 \rangle/N$.
The Polyakov loop is the order parameter of deconfinement transition.
In the condensed phase, the Polyakov loop is nonzero and colored particles are deconfined.

\begin{figure}[h]
\begin{center}
 \includegraphics[width=.6\textwidth]{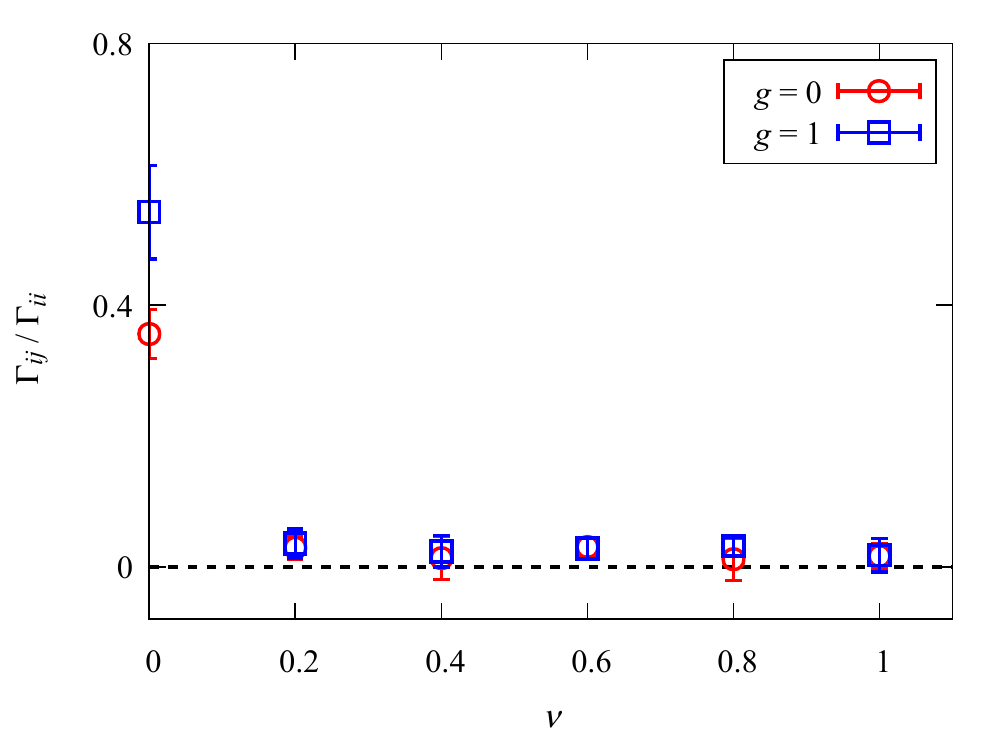}
\caption{
\label{figCN}
Ratio of the off-diagonal component $\Gamma_{ij}$ $(i\neq j)$ to the diagonal component $\Gamma_{ii}$ characterizing the color-flavor locking.
The tachyonic mass is $m=2$.
}
\end{center}
\end{figure}

The scalar coupling constant $\nu$ controls the color-flavor locking.
In the color-flavor locked phase, the vacuum is given by the diagonal color-flavor matrix
\begin{equation}
\label{eqCFL1}
 \phi_{ai} \propto \diag(1, \cdots ,1)
\end{equation}
up to gauge transformation.
We define the gauge invariant operator
\begin{equation}
 G_{ij}(x) = \phi^\dagger_i(x) \phi_j(x)
.
\end{equation}
When $\phi_i$ is given by Eq.~\eqref{eqCFL1}, $G_{ij}$ is diagonal in flavor space.
Note however that the expectation value $\langle G_{ij}\rangle$ is trivially diagonal because the action \eqref{eqSlat} and the diagonal component $G_{ii}$ are even under the inversion $\phi_i \leftrightarrow -\phi_i$ while the off-diagonal component $G_{ij}$ $(i\neq j)$ is odd.
To obtain meaningful results, we calculated the long-range limit of the two-point function
\begin{equation}
\label{eqGamma}
 \Gamma_{ij} = \left\{ \lim_{|x-y| \to \infty} \langle G^\dagger_{ij}(x) G_{ij}(y) \rangle \right\}^{1/2}
,
\end{equation}
where $i$ and $j$ are not contracted.
The diagonality of $\Gamma_{ij}$ characterizes the color-flavor locking.
As shown in Fig.~\ref{figCN}, the off-diagonal component of $\Gamma_{ij}$ is zero in $\nu>0$.
Therefore the color-flavor locking is realized.

\section{Non-Abelian vortex}

We introduce external U(1) magnetic fields to generate non-Abelian vortices.
We consider a homogeneous magnetic field $-B$ in the $x_3$ direction.
The U(1) link variables are set to $V_\mu = \exp(ieA^0_\mu T^0)$ with $A^0_1 = B x_2$, $A^0_2 = -B x_1$, and $A^0_3 = 0$.
The parameters are fixed at $m=2$ and $\lambda=\nu=1$ to realize the color-flavor locked phase.
The boundary conditions in the $x_1$ and $x_2$ directions are not periodic but the Neumann-like one where the covariant derivatives in Eq.~\eqref{eqScon} (or the differences in Eq.~\eqref{eqSlat}) perpendicular to these boundaries are set to be zero.
There are two reasons; to change magnetic fields finely and to break translational symmetry explicitly, as explained later.
The boundary condition in the $x_3$ direction is periodic.
Since translational invariance is violated in the $x_1$ and $x_2$ directions, the long-range limit in Eqs.~\eqref{eqDelta} and \eqref{eqGamma} is taken in the $x_3$ direction.
Other setups are the same as the above calculation without external magnetic fields.

A vortex is defined by the counter integral of a phase.
One would naively calculate the phase of the scalar field $\phi_i \propto e^{i\theta}$, but it is not gauge invariant.
We calculated the phase of the gauge invariant operator \eqref{eqH}.
We define the phase difference between neighboring sites
\begin{equation}
\label{eqdeltaH}
 \delta_\mu \theta(x) = \arg \left\{ {H^\dagger(x) H(x+\hat{\mu})} \right\}
,
\end{equation}
the vortex density of a plaquette
\begin{equation}
\label{eqqx}
 q(x) = \frac{1}{2\pi} \{ \delta_1 \theta(x) + \delta_2\theta(x+\hat{1})- \delta_1\theta(x+\hat{2}) - \delta_2\theta(x) \}
,
\end{equation}
and the total vortex number
\begin{equation}
\label{eqQ}
 Q = \sum_{x_1,x_2} q(x)
\end{equation}
in the $x_1$-$x_2$ plane.
The classical value $Q$ is an integer by definition.
However, the expectation value $\langle Q \rangle$ can be a non-integer because all vortex states are mixed by quantum fluctuation.
In terms of the Monte Carlo simulation, $Q$ is an integer in each configuration of the Monte Carlo ensemble.
When the configurations with different $Q$ exist in one ensemble, the ensemble average $\langle Q \rangle$ is a non-integer.
Such a non-integer value is characteristic in full quantum calculation, which is not obtained in mean-field calculation.

\begin{figure}[h]
\begin{center}
 \includegraphics[width=.6\textwidth]{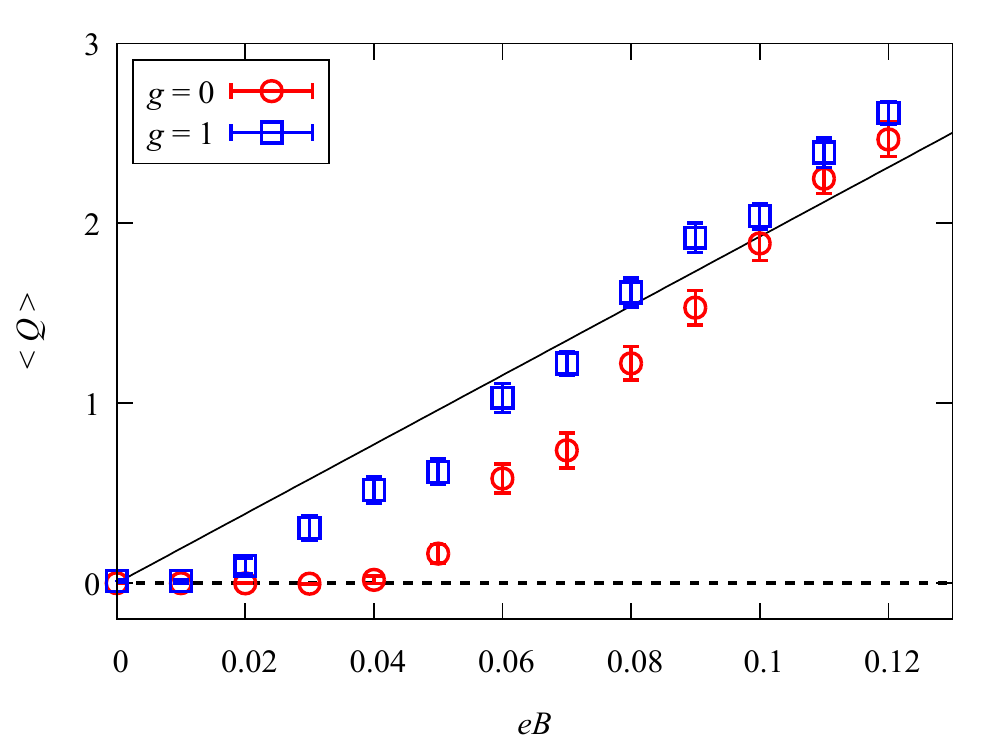}
\caption{
\label{figQ}
Vortex number $\langle Q \rangle$.
The solid line is the total external magnetic flux $\int eB d^2x / 2\pi$.
}
\end{center}
\end{figure}

The expectation value $\langle Q \rangle$ is shown in Fig.~\ref{figQ}.
Non-Abelian vortices are generated above the critical magnetic field $eB \simeq 0.04$ at $g=0$ and $eB \simeq 0.02$ at $g=1$.
The critical magnetic field is larger at $g=0$ because the condensate is larger as shown in Fig.~\ref{figCM}.
The total flux of the external magnetic field $\int eB d^2x /2\pi$ is also drawn in Fig.~\ref{figQ}.
The vortex number is consistent with the total magnetic flux in strong magnetic fields.
(If boundary conditions are periodic, the total magnetic flux must be quantized to integer values, $\int eB d^2x / 2\pi =1,2,\cdots$, because of the Stokes theorem in a torus.
We can calculate few data points in Fig.~\ref{figQ}.
This is the first reason to impose non-periodic boundary conditions.)

\begin{figure}[h]
\begin{center}
 \includegraphics[width=.49\textwidth]{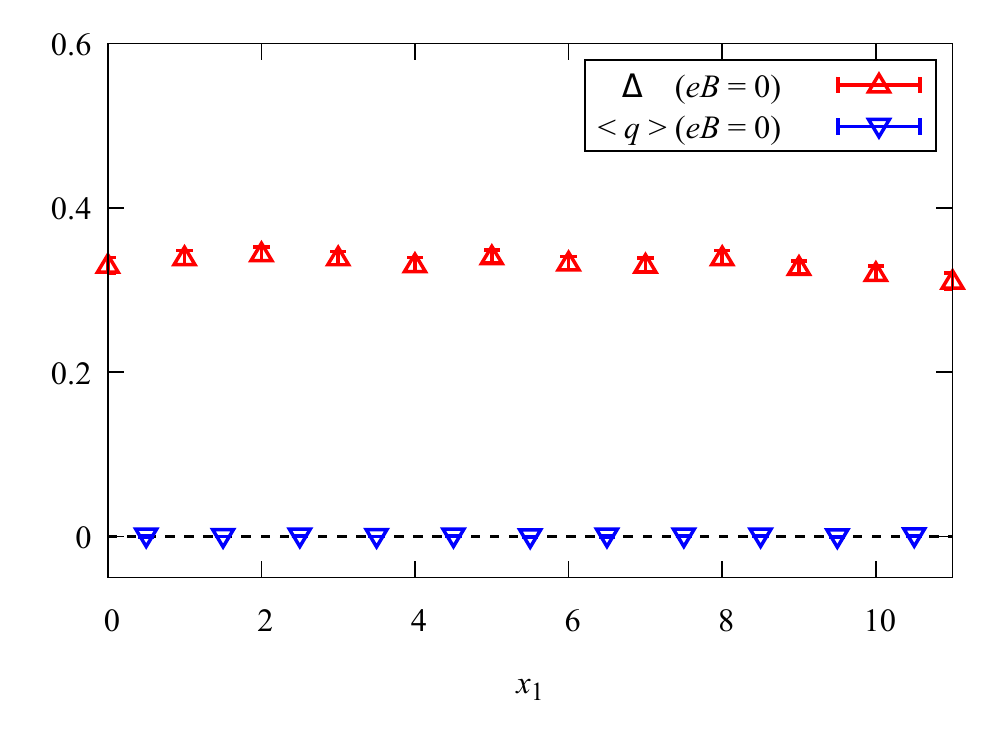}
 \includegraphics[width=.49\textwidth]{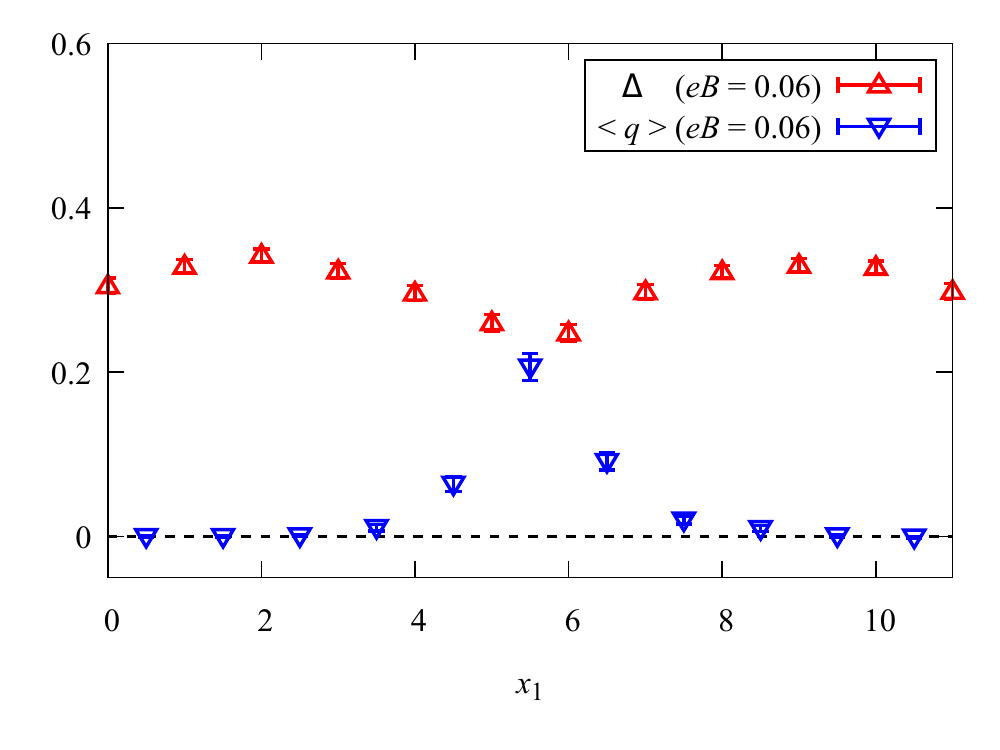}
\caption{
\label{figX}
Coordinate dependence of the vortex density $\langle q(x) \rangle$ and the condensate $\Delta(x)$ at $eB=0$ (left) and $eB=0.06$ (right).
The perpendicular coordinate is fixed at $x_2=L/2$.
The gauge coupling constant is $g=1$.
}
\end{center}
\end{figure}

In Fig.~\ref{figX}, we plot the coordinate dependence of the vortex density $\langle q(x) \rangle$ and the condensate $\Delta(x)$.
At $eB=0$, the vortex density is zero and the condensate is homogeneous.
At $eB=0.06$, nonzero vortex density is induced.
The vortex density is inhomogeneous and favors the center of the box.
The condensate is suppressed inside the vortex.
This is for the same reason as rotational vortices, which favor the center to minimize rotational energy \cite{Hayata:2014kra}.
(If boundary conditions are periodic, the system is translationally invariant.
The coordinate dependence cannot be observed.
This is the second reason to impose non-periodic boundary conditions.)

\begin{figure}[t]
\begin{center}
 \includegraphics[width=.98\textwidth]{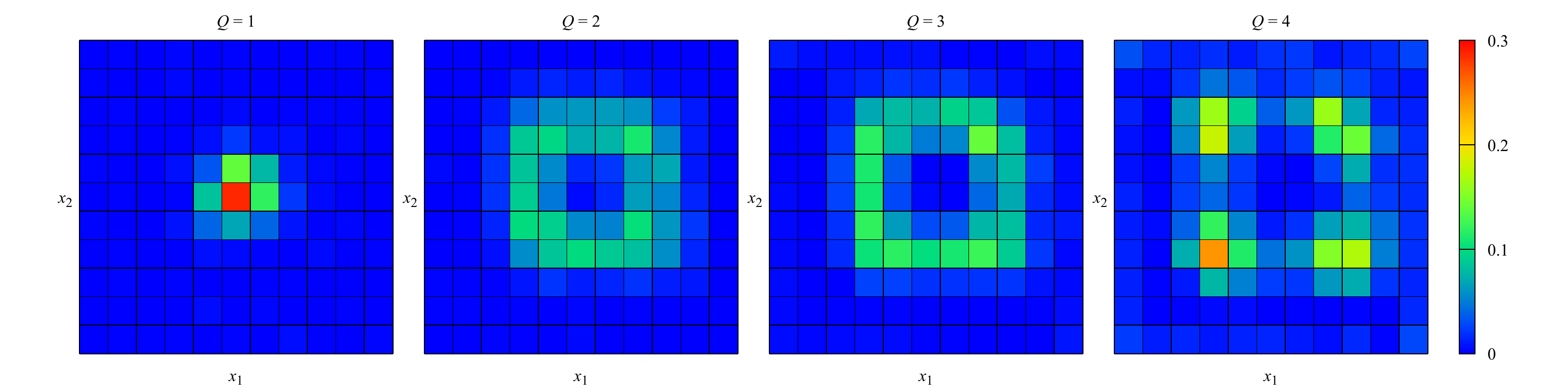}
\caption{
\label{figQS}
Spatial distribution of the vortex density $\langle q(x)\rangle_Q$ with fixed vortex numbers.
The data of $\langle q(x)\rangle_{Q=1}$ are obtained at $eB=0.06$ and the data of $\langle q(x)\rangle_{Q=2,3,4}$ are obtained at $eB=0.12$.
The gauge coupling constant is $g=1$.
}
\end{center}
\end{figure}

Although the vortex number is $\langle Q \rangle \simeq 1$ at $eB=0.06$, Fig.~\ref{figX} is not exactly the spatial distribution of one non-Abelian vortex.
The configurations with different vortex numbers are mixed.
The ensemble at $eB=0.06$ consists of the configurations with $Q=-1$ (1\%), $Q=0$ (17\%), $Q=1$ (64\%), and $Q=2$ (18\%).
We can separately analyze each vortex number sector.
We sorted the configurations by the vortex number, and calculated the expectation value $\langle \cdots \rangle_Q$ in the ensemble with the vortex number $Q$.
The vortex density $\langle q(x) \rangle_{Q=1}$ is shown in the left panel of Fig.~\ref{figQS}.
This is exactly the spatial distribution of one non-Abelian vortex.
The vortex is localized at the center of the lattice box.

Multi-vortex distributions can be obtained in the same manner.
At $eB=0.12$, the ensemble consists of the configurations with $Q=0$ (1\%), $Q=1$ (8\%), $Q=2$ (29\%), $Q=3$ (47\%), and $Q=4$ (15\%).
The results of $Q=2$, 3, and 4 are shown in Fig.~\ref{figQS}.
At $Q=2$ and 3, we see a donut-like shape.
In each configuration, the distribution is not a donut but two or three separate vortices.
Since the lattice box has $\pi/2$-rotational symmetry, the expectation value is given by the superposition of the rotated distributions of two or three vortices, and then results in a rotationally symmetric donut.
The center hole in the donut comes from nonzero separation between vortices.
At $Q=4$, we clearly see four vortices.
This is because the four-vortex distribution is a priori symmetric under the $\pi/2$ rotation.
These multi-vortex distributions suggest the repulsion between non-Abelian vortices.
In the normalization of Eqs.~\eqref{eqdeltaH} to \eqref{eqQ}, the vortex number of one Abelian vortex is $Q=3$.
Due to the repulsion, one Abelian vortex with $Q=3$ will be unstable and decay into three non-Abelian vortices with $Q=1$.
The repulsion and instability are consistent with mean-field analysis \cite{Nakano:2007dr,Alford:2016dco}.

There are two remarks:
The first one is about volume scaling.
The obtained results are the superpositions of all vortex states with finite action.
Some vortices, known as global vortices, have logarithmically divergent action in the infinite volume limit \cite{Nitta:2007dp,Eto:2009wu}.
The contributions from such vortices will be suppressed in larger volumes.
The volume scaling of the above results is a nontrivial question.
The second one is about the operator to count the vortex number.
It is not unique.
For example, another definition is $\delta_\mu \theta \equiv \arg\{\phi^\dagger_i(x) U_\mu(x) \phi_i(x+\hat{\mu}) \}$.
Since this definition is gauge invariant for each flavor, we can calculate the vortex number of each flavor independently.
In this definition, however, the vortex number is not necessarily an integer because the phase difference is defined by the inner product of vectors.
We numerically checked that the total vortex number in this definition is almost consistent with Fig.~\ref{figQ}.

\section{Summary and perspective}

We studied non-Abelian vortices by the Monte Carlo simulation of the non-Abelian Higgs model.
We applied external magnetic fields to excite non-Abelian vortices.
The similar analysis will be possible by rotating the lattice \cite{Yamamoto:2013zwa}.
We confirmed the repulsive vortex-vortex interaction from the spatial distribution in Fig.~\ref{figQS}.
More quantitative analysis will be possible by calculating the vortex-vortex potential from the 't Hooft loop \cite{Hoelbling:2000su,DelDebbio:2000cx,deForcrand:2000fi}.
Our ultimate goal is the lattice simulation of vortices in high-density QCD.
This is now impossible due to the fermion sign problem.
Approximate study might be possible in sign-problem-free fermion models with nonzero density, e.g., the generalization of two-color or isospin or chiral density.

\ack
The author was supported by JSPS KAKENHI Grant Number 15K17624. 
The numerical simulations were carried out on SX-ACE in Osaka University.

\bibliographystyle{ptephy}
\bibliography{paper}

\begin{thebibliography}{10}

\bibitem{1991qvhi.book.....D}
R.~J. {Donnelly},
\newblock {\em {Quantized Vortices in Helium II}},
\newblock  (Cambridge University Press, 1991).

\bibitem{RevModPhys.82.109}
B.~Rosenstein and D.~Li, Rev. Mod. Phys., {\bf 82}, 109 (2010).

\bibitem{RevModPhys.81.647}
A.~L. Fetter, Rev. Mod. Phys., {\bf 81}, 647 (2009).

\bibitem{Greensite:2011zz}
J.~Greensite, Lect. Notes Phys., {\bf 821}, 1 (2011).

\bibitem{Tong:2005un}
D.~Tong (),  {{arXiv:hep-th/0509216}}.

\bibitem{Eto:2006pg}
M.~Eto, Y.~Isozumi, M.~Nitta, K.~Ohashi, and N.~Sakai, J. Phys., {\bf A39},
  R315 (2006),  {{arXiv:hep-th/0602170}}.

\bibitem{Shifman:2007ce}
M.~Shifman and A.~Yung, Rev. Mod. Phys., {\bf 79}, 1139 (2007),
  {{arXiv:hep-th/0703267}}.

\bibitem{Eto:2013hoa}
M.~Eto, Y.~Hirono, M.~Nitta, and S.~Yasui, PTEP, {\bf 2014}, 012D01 (2014),
  {{arXiv:1308.1535}}.

\bibitem{Balachandran:2005ev}
A.~P. Balachandran, S.~Digal, and T.~Matsuura, Phys. Rev. D, {\bf 73}, 074009
  (2006),  {{arXiv:hep-ph/0509276}}.

\bibitem{Chavel:1996hh}
M.~Chavel, Phys. Lett. B, {\bf 378}, 227 (1996),  {{arXiv:hep-lat/9603005}}.

\bibitem{Kajantie:1998bg}
K.~Kajantie, M.~Karjalainen, M.~Laine, J.~Peisa, and A.~Rajantie, Phys. Lett.
  B, {\bf 428}, 334 (1998),  {{arXiv:hep-ph/9803367}}.

\bibitem{Chernodub:1998kj}
M.~N. Chernodub, M.~I. Polikarpov, A.~I. Veselov, and M.~A. Zubkov, Phys. Lett.
  B, {\bf 432}, 182 (1998),  {{arXiv:hep-lat/9804002}}.

\bibitem{Kajantie:1998zn}
K.~Kajantie, M.~Laine, T.~Neuhaus, J.~Peisa, A.~Rajantie, and K.~Rummukainen,
  Nucl. Phys., {\bf B546}, 351 (1999),  {{arXiv:hep-ph/9809334}}.

\bibitem{Kajantie:1999ih}
K.~Kajantie, M.~Laine, T.~Neuhaus, A.~Rajantie, and K.~Rummukainen, Nucl.
  Phys., {\bf B559}, 395 (1999),  {{arXiv:hep-lat/9906028}}.

\bibitem{Kajantie:2000cw}
K.~Kajantie, M.~Laine, T.~Neuhaus, A.~Rajantie, and K.~Rummukainen, Phys. Lett.
  B, {\bf 482}, 114 (2000),  {{arXiv:hep-lat/0003020}}.

\bibitem{Davis:2000kv}
A.~C. Davis, T.~W.~B. Kibble, A.~Rajantie, and H.~Shanahan, JHEP, {\bf 11}, 010
  (2000),  {{arXiv:hep-lat/0009037}}.

\bibitem{Chernodub:2005be}
M.~N. Chernodub, R.~Feldmann, E.~M. Ilgenfritz, and A.~Schiller, Phys. Rev. D,
  {\bf 71}, 074502 (2005),  {{arXiv:hep-lat/0502009}}.

\bibitem{Wenzel:2007uh}
S.~Wenzel, E.~Bittner, W.~Janke, and A.~M.~J. Schakel, Nucl. Phys., {\bf B793},
  344 (2008),  {{arXiv:0708.0903}}.

\bibitem{MacKenzie:2007ps}
R.~MacKenzie, F.~Nebia-Rahal, and M.~B. Paranjape, Phys. Rev. D, {\bf 81},
  114505 (2010),  {{arXiv:0710.3236}}.

\bibitem{Bietenholz:2012ud}
W.~Bietenholz, M.~Bogli, F.~Niedermayer, M.~Pepe, F.~G. Rejon-Barrera, and
  U.~J. Wiese, JHEP, {\bf 03}, 141 (2013),  {{arXiv:1212.0579}}.

\bibitem{PhysRevLett.75.4642}
G.~Ortiz and D.~M. Ceperley, Phys. Rev. Lett., {\bf 75}, 4642 (1995).

\bibitem{PhysRevB.72.094511}
E.~Bittner, A.~Krinner, and W.~Janke, Phys. Rev. B, {\bf 72}, 094511 (2005).

\bibitem{PhysRevB.89.224516}
D.~E. Galli, L.~Reatto, and M.~Rossi, Phys. Rev. B, {\bf 89}, 224516 (2014).

\bibitem{PhysRevA.95.053603}
L.~Madeira, S.~Gandolfi, and K.~E. Schmidt, Phys. Rev. A, {\bf 95}, 053603
  (2017).

\bibitem{Iida:2000ha}
K.~Iida and G.~Baym, Phys. Rev. D, {\bf 63}, 074018 (2001),
  {{arXiv:hep-ph/0011229}}.

\bibitem{Iida:2001pg}
K.~Iida and G.~Baym, Phys. Rev. D, {\bf 65}, 014022 (2001),
  {{arXiv:hep-ph/0108149}}.

\bibitem{Giannakis:2001wz}
I.~Giannakis and H.-c. Ren, Phys. Rev. D, {\bf 65}, 054017 (2002),
  {{arXiv:hep-ph/0108256}}.

\bibitem{Alford:1998mk}
M.~G. Alford, K.~Rajagopal, and F.~Wilczek, Nucl. Phys., {\bf B537}, 443
  (1999),  {{arXiv:hep-ph/9804403}}.

\bibitem{Montvay:1994cy}
I.~Montvay and G.~Munster,
\newblock {\em {Quantum fields on a lattice}},
\newblock  (Cambridge University Press, 1997).

\bibitem{Alford:2018mqj}
M.~G. Alford, G.~Baym, K.~Fukushima, T.~Hatsuda, and M.~Tachibana (),
  {{arXiv:1803.05115}}.

\bibitem{Hayata:2014kra}
T.~Hayata and A.~Yamamoto, Phys. Rev. A, {\bf 92}, 043628 (2015),
  {{arXiv:1411.5195}}.

\bibitem{Nakano:2007dr}
E.~Nakano, M.~Nitta, and T.~Matsuura, Phys. Rev. D, {\bf 78}, 045002 (2008),
  {{arXiv:0708.4096}}.

\bibitem{Alford:2016dco}
M.~G. Alford, S.~K. Mallavarapu, T.~Vachaspati, and A.~Windisch, Phys. Rev. C,
  {\bf 93}, 045801 (2016),  {{arXiv:1601.04656}}.

\bibitem{Nitta:2007dp}
M.~Nitta and N.~Shiiki, Phys. Lett., {\bf B658}, 143 (2008),
  {{arXiv:0708.4091}}.

\bibitem{Eto:2009wu}
M.~Eto, E.~Nakano, and M.~Nitta, Nucl. Phys., {\bf B821}, 129 (2009),
  {{arXiv:0903.1528}}.

\bibitem{Yamamoto:2013zwa}
A.~Yamamoto and Y.~Hirono, Phys. Rev. Lett., {\bf 111}, 081601 (2013),
  {{arXiv:1303.6292}}.

\bibitem{Hoelbling:2000su}
C.~Hoelbling, C.~Rebbi, and V.~A. Rubakov, Phys. Rev. D, {\bf 63}, 034506
  (2001),  {{arXiv:hep-lat/0003010}}.

\bibitem{DelDebbio:2000cx}
L.~Del~Debbio, A.~Di~Giacomo, and B.~Lucini, Nucl. Phys., {\bf B594}, 287
  (2001),  {{arXiv:hep-lat/0006028}}.

\bibitem{deForcrand:2000fi}
P.~de~Forcrand, M.~D'Elia, and M.~Pepe, Phys. Rev. Lett., {\bf 86}, 1438
  (2001),  {{arXiv:hep-lat/0007034}}.

\end{thebibliography}

\end{document}